\RequirePackage{lineno}
\documentclass[twocolumn,prl,amsmath,amssymb,preprintnumbers,showkeys,showpacs,superscriptaddress,floatfix,nofootinbib,aps]{revtex4-1}

\def\apj{ApJ}

\def\apss{Ap\&SS}               
\def\planss{P\&SS}              

\def\jgr{Journal of Geophysical Research}

\usepackage{graphicx,color}
\usepackage{multirow}
\graphicspath{{Plots/}}

\usepackage{url}

\def\units#1{\hbox{$\,{\rm #1}$}}
\def\degrees{\hbox{$^\circ$}}

\newcommand{\HelioProp}{{\sc HelioProp}}
\newcommand{\Dragon}{{\sc Dragon}}

\begin{document}


\title{Inferring the local interstellar spectrum of cosmic ray protons from PAMELA data}

\author{F.~Loparco}
\email{loparco@ba.infn.it}
\affiliation{Istituto Nazionale di Fisica Nucleare, Sezione di Bari, 70126 Bari, Italy}
\affiliation{Dipartimento di Fisica ``M. Merlin" dell'Universit\`a e del Politecnico di Bari, I-70126 Bari, Italy}

\author{L.~Maccione}
\email{luca.maccione@lmu.de}
\affiliation{Ludwig-Maximilians-Universit\"{a}t, Theresienstra\ss{}e 37, D-80333 M\"{u}nchen, Germany}
\affiliation{Max-Planck-Institut f\"{u}r Physik (Werner Heisenberg Institut), F\"{o}hringer Ring 6, D-80805 M\"{u}nchen, Germany}

\author{M.~N.~Mazziotta}
\email{mazziotta@ba.infn.it} 
\affiliation{Istituto Nazionale di Fisica Nucleare, Sezione di Bari, 70126 Bari, Italy}

\begin{abstract}
We take advantage of the cosmic-ray (CR) proton data collected by the PAMELA detector in several campaigns covering the period 2006-2009 to derive the local interstellar CR proton spectrum (LIS), outside the solar system. We describe the propagation of CR protons in the solar system by means of a detailed model of the charge-sign dependent drifts occurring while CRs diffuse in the irregular solar magnetic field. We fit PAMELA data for each year of data taking and find that they are well described by a unique time-independent LIS. We then discuss the consequences of this LIS on galactic propagation models. We find that diffusive reacceleration is strongly constrained, thus confirming previous results derived from different channels.
\end{abstract}



\preprint{LMU-ASC 36/13, MPP-2013-142}

\maketitle

{\em Introduction ---} The propagation of Cosmic Rays (CRs) is far from being understood. Below a few 10 GeV several different effects concur to modify the energy spectrum injected by their sources. Besides diffusion, also reacceleration in the turbulent galactic plasma and convection in stellar winds can play a significant role. Upon parameterizing the effects of these physical processes in a convenient way, existing semi-analytic \cite{Usineweb} and numerical \cite{Galpropweb,Dragonweb} propagation codes are able to compute the energy spectrum of the propagated CRs at the solar system location. 

However, this cannot be directly compared with the spectrum observed by terrestrial experiments. Indeed, the CR spectra that we observe at Earth are different from the local interstellar spectra (LIS) that are computed in galactic propagation models, because of energy losses during propagation in the solar system \cite{1965P&SS...13....9P,Gleeson_1968ApJ}, that affect the CR spectrum below a few GeV \cite{1996ApJ...464..507C} and vary on a yearly basis according to the evolution of the Sun. The presence of these effects and the extreme difficulty of describing them in sufficient detail, have hampered the possibility to understand the LIS.

Recently, two major experimental breakthroughs have radically changed the situation. First, the analyses of the $\gamma$-ray spectra from individual molecular clouds \cite{Neronov:2011wi} and of the diffuse galactic $\gamma$-ray emission \cite{Dermer:2013iwa,2012AIPC.1505...37C} have allowed us to infer with some confidence the interstellar CR proton spectrum in the local environment. Second, the PAMELA Collaboration has provided direct measurements of the CR proton spectra for the years 2006, 2007, 2008 and 2009 \cite{Adriani:2013as}, that will consent to analyze the effects of solar propagation and to possibly disentangle them from the ones of galactic propagation.

Inferring the proton LIS from $\gamma$-ray data requires the knowledge of the cross section for $\gamma$-ray production in $pp$ collisions and, if the diffuse data are used, of the distribution of hydrogen gas in the galaxy. In addition, the contributions of CR Helium emission and of bremsstrahlung and inverse Compton emissions by CR electrons and positrons, although subdominant, need to be accounted for. All these quantities are subject to large uncertainties, that make this procedure burdened by severe systematics. 
The question whether the LIS inferred from Fermi-LAT data and PAMELA spectra are compatible is therefore crucial in order to validate the Fermi-LAT analysis and to reduce systematic uncertainties by fully exploiting the multi-messenger wealth of data available now.

In this Letter we derive the proton LIS from PAMELA data. By assuming a simple shape for the LIS and by accurately sampling the parameter space of solar propagation, we calculate the best fitting LIS and solar parameters for each year of data taking. We show that PAMELA data can be simultaneously fit using one single LIS, independent of the data-taking period. This requires to slightly adjust the solar parameters in each year, as expected due to their natural time dependence. A previous analysis of the PAMELA spectra focused mainly on the properties of solar propagation and fixed the LIS {\it a priori} \cite{Potgieter:2013cwj}. We assume instead some parametric analytical LIS models and simulate propagation in the solar system using the dedicated code \HelioProp~\cite{Maccione:2012cu}, that includes an accurate description of the solar modulation physics. 

Once a shape for the LIS is given, the natural question arises, whether it can be reproduced within a model of galactic propagation. We investigate also this problem and show that the proton LIS is compatible with a single power-law injection spectrum, as expected in the popular diffusive shock acceleration models \cite{Caprioli:2011ze}. Furthermore, we place limits on the strength of diffusive reacceleration and of convection in the interstellar medium.

{\em Method ---} In order to simplify the problem and to make the comparison with Fermi-LAT spectrum easier, we consider 2 different possibilities for the shape of the LIS spectrum $J_{\rm LIS}(p)$:
\begin{itemize}
\item single power-law (SPL) $J_{\rm LIS}(p) = k_{0}\beta(p/p_{0})^{-\alpha}$, with $p_{0}=1~\units{GeV}$
\item broken power-law (BPL) $J_{\rm LIS}(p) = k_{0}\beta(p/p_{\rm br})^{-\alpha_{1}}$ if $p<p_{\rm br}$ and $J_{\rm LIS}(p) = k_{0}\beta(p/p_{\rm br})^{-\alpha_{2}}$ if $p\geq p_{\rm br}$ 
\end{itemize}
where $p$ is the CR momentum and $\beta$ is the velocity of the CR in units of the speed of light $c=1$.
Units will be $\units{GeV^{-1} m^{-2} s^{-1} sr^{-1}}$ for $k_{0}$ and $\units{GeV}$ for momenta unless otherwise stated. The conversion from this spectrum to a spectrum in kinetic energy $T$ can be easily achieved by adding a factor of $dp/dT = 1/\beta$.

We describe solar modulation with the \HelioProp\ numerical code which solves the CR transport equation in the heliosphere accounting for charge-dependent effects \cite{Maccione:2012cu}. Because of drifts in the large scale gradients of the solar magnetic field (SMF), the modulation effect depends on the particle charge, including its sign \cite{1996ApJ...464..507C}. Therefore, it depends on the polarity of the SMF, which changes periodically every $\sim$11 years \cite{wilcox}. 

The SMF has also opposite polarities in the northern and southern hemispheres:~at the interface between opposite polarity regions, a heliospheric current sheet (HCS) is formed (see e.g.~\cite{1981JGR....86.8893B}). The HCS swings then in a region whose angular extension is described phenomenologically by a tilt angle, whose magnitude depends on solar activity. Since particles crossing the HCS suffer from additional drifts because of the different orientation of the magnetic field lines, the intensity of the modulation depends on the extension of the HCS. The geometry of drifts is such that particles having the same charge sign as the polarity of the magnetic field arrive to Earth preferentially from the poles, and particles having opposite charge sign mainly drift inwards along the HCS \cite{2012Ap&SS.339..223S}. This results into different energy losses of particles with opposite charge signs.

Besides the magnitude of the tilt angle, another very important parameter of the model is related to diffusion. 
We assume that diffusion is anisotropic, with the parallel diffusion coefficient $D(\rho)=\lambda(\rho) v/3$, with $\lambda = \lambda_{0}(\rho/1~\units{GV})^{\delta_{\odot}}$ being the momentum-dependent mean-free-path, and the perpendicular diffusion coefficients being proportional to $D$ through the constant $k_{\perp}=0.02$ \cite{Maccione:2012cu}. For the solar magnetic field we will assume a simple Parker spiral. We will fix the values of the tilt angle and of the solar magnetic field at the Earth position $B_{0}$ to the observed ones for each year of data taking. 
According to \cite{Potgieter:2013cwj} we will take for the tilt angle $15.7\degrees$, $14.0\degrees$, $14.3\degrees$, $10.0\degrees$, while for $B_{0} ({\rm nT})$ we use 5.05, 4.50, 4.25, 3.94 for the years 2006, 2007, 2008 and 2009 respectively. We will instead take the very uncertain $\lambda_{0}$ and $\delta_{\odot}$ as free parameters and fit them along with the proton LIS. 


The proton spectra at Earth are evaluated after folding the appropriate LIS spectrum with the solar propagation code \HelioProp. The PAMELA data are then fitted to the computed spectrum at Earth.
The fit is performed in the range of kinetic energy $0.085\div175~\units{GeV}$ using the MINUIT package implemented in the ROOT framework~\cite{ROOT}. 
The data up to $48~\units{GeV}$ are taken from Table 1 of  Ref.~\cite{Adriani:2013as}, while the data from 
$48~\units{GeV}$ to $1~\units{TeV}$ are taken from Ref.~\cite{Adriani:2011cu}. 
The error bars shown in the plots are evaluated by adding in quadrature statistical and systematic uncertainties. 

Finally we study the consequences of the fitted LIS on galactic propagation parameters. The CR transport in the Galaxy is described by the well-known diffusion-convection-reacceleration-energy-loss equation \cite{1964ocr..book.....G,1965P&SS...13....9P}. For each CR particle, we solve the set of coupled transport equations with the numerical code \Dragon\ \cite{Dragonweb,Evoli:2008dv,DiBernardo:2010is} in its 2D version. For our purposes, we can take the diffusion coefficient as a scalar with the following dependences on the rigidity $\rho$: $D_{{\rm gal}}(\rho) = \beta^{\eta}~ D_0 \left( \rho/\rho_0 \right)^\delta$, with $\rho_{0}=3~\units{GV}$ and $\eta$ and $\delta$ being constants \cite{DiBernardo:2010is}. Reacceleration is described as diffusion in momentum space, with the diffusion coefficient $D_{pp}\propto p^{2} v_{\rm A}^{2}/D_{\rm gal}$ \cite{Berezinsky_book}, with $v_{\rm A}$ being the Alfv\'en velocity. For convection, we assume a convective velocity $v_{C} \propto |z|$ directed outwards the galactic plane, as it is usually done.

\begin{figure}[tbp]
\begin{center}
\includegraphics[width=1\columnwidth,keepaspectratio,clip]{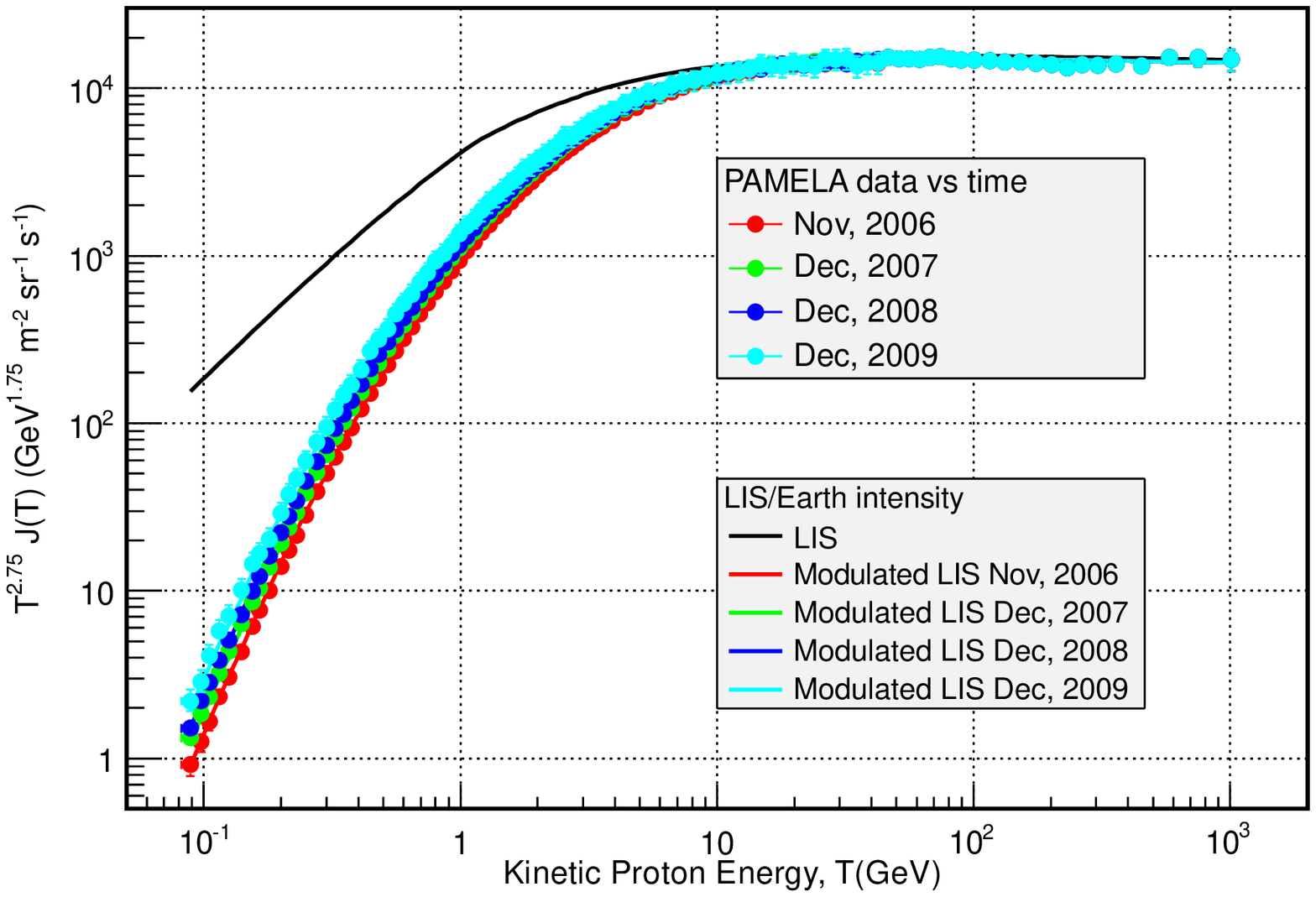}
\includegraphics[width=1\columnwidth,keepaspectratio,clip]{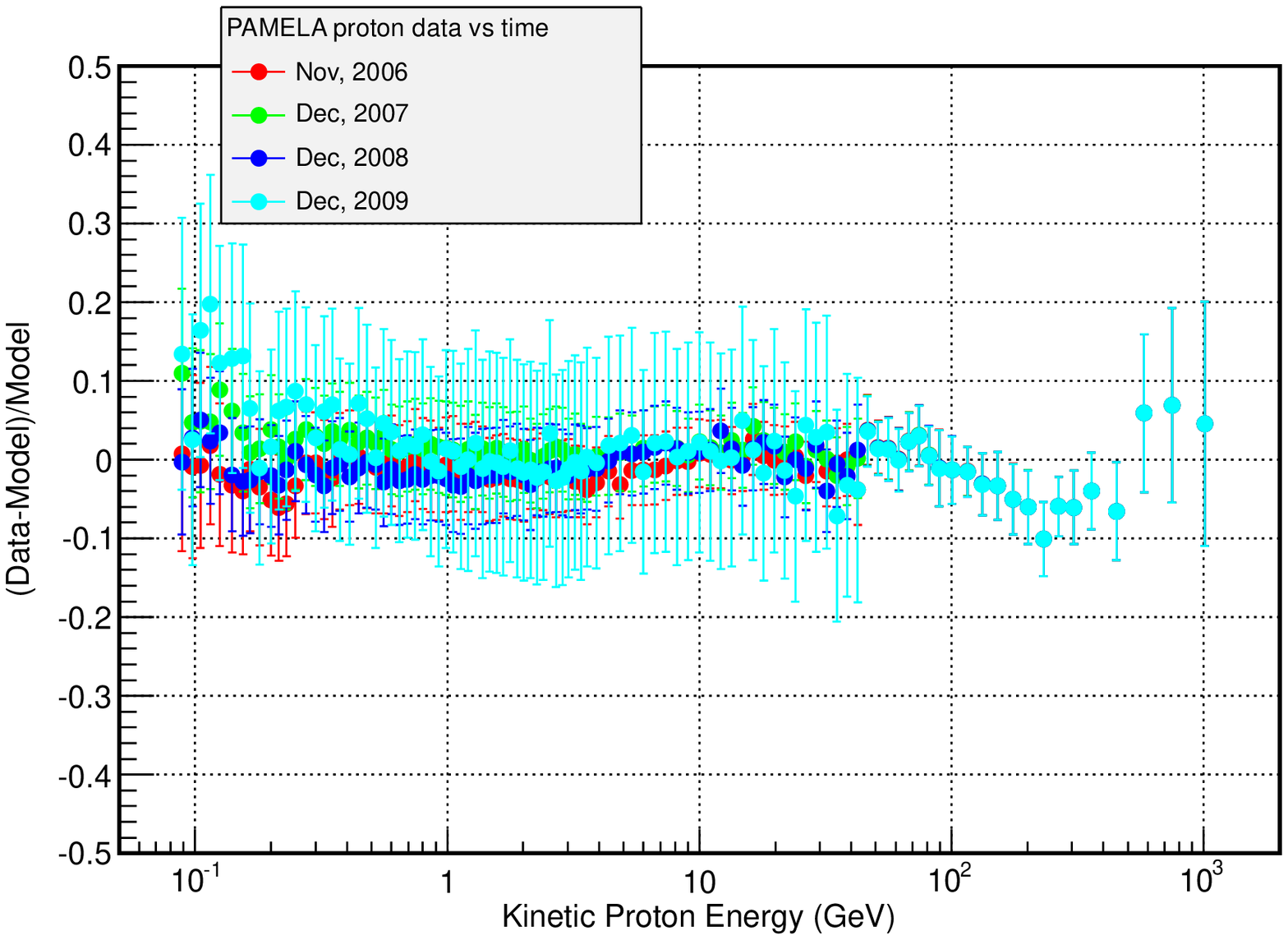}
\end{center}
\caption{Joint fit of the PAMELA proton data collected in November 2006 (red), December 2007 (green), December 2008 (blue)
and December 2009 (cyan) with a simple momentum power law LIS folded with solar modulation. The top panel shows the fit results superimposed to the the data points. The black line shows the corresponding LIS, while the colored lines show the fitted spectra at the Earth. 
The bottom panel shows the fit residuals.}
\label{Fig3}
\end{figure}

\begin{figure}[tbp]
\begin{center}
\includegraphics[width=1\columnwidth,keepaspectratio,clip]{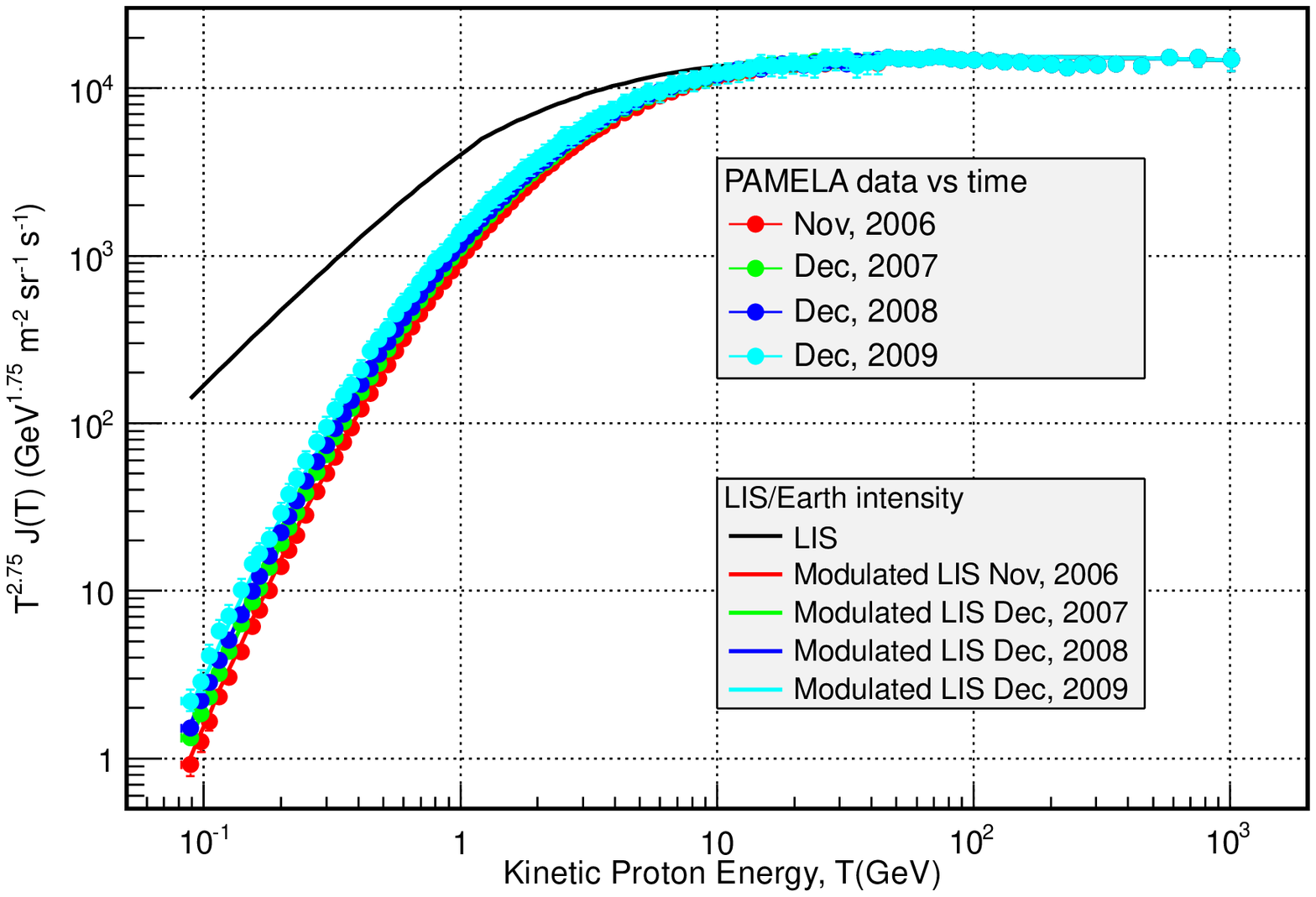}
\includegraphics[width=1\columnwidth,keepaspectratio,clip]{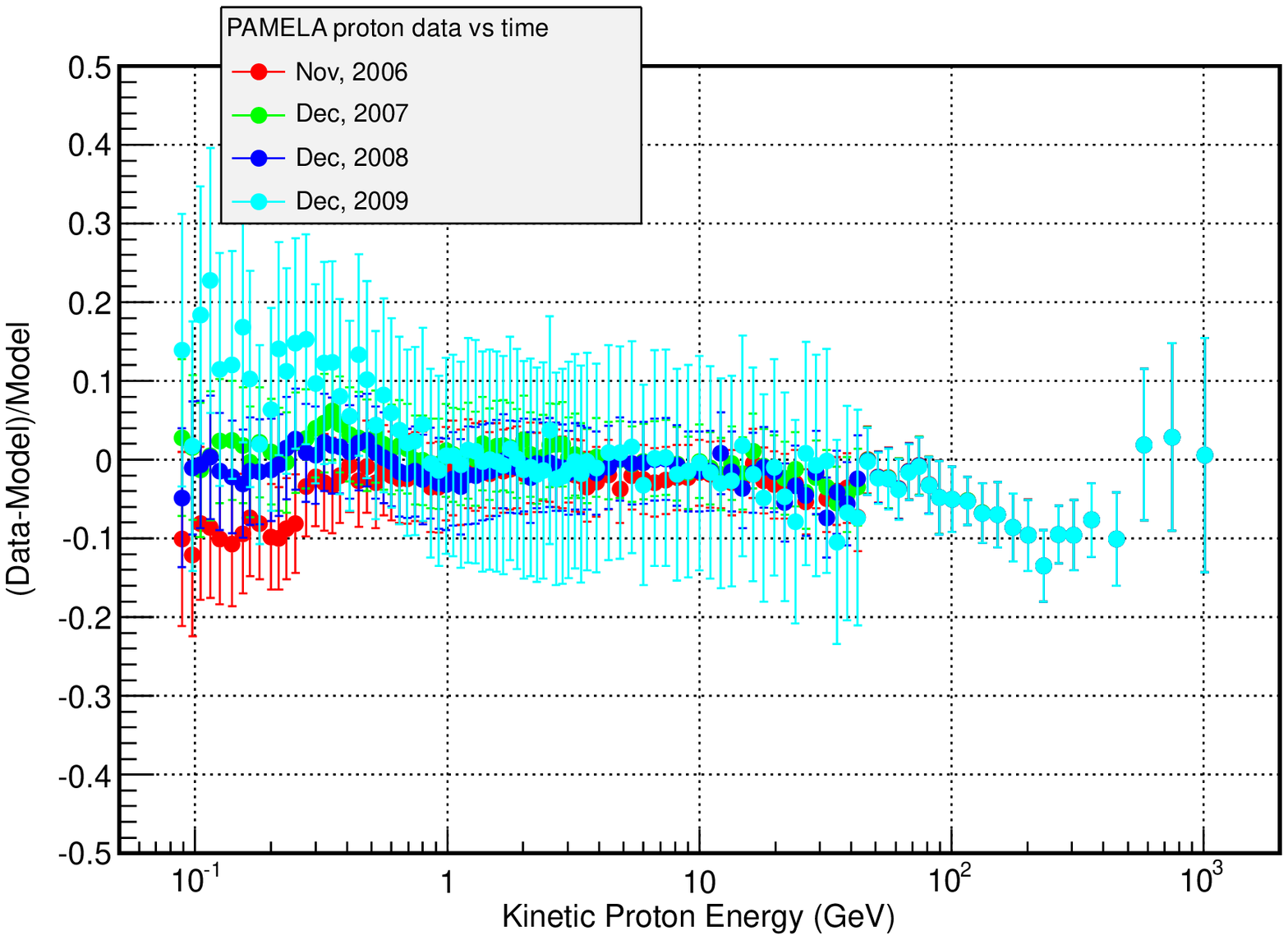}
\end{center}
\caption{Same as Fig.~\ref{Fig3} but for the broken power law LIS.}
\label{Fig4}
\end{figure}

{\em Results --- } We show in Tables \ref{tab:fitSP} and \ref{tab:fitBP} the values of the model parameters with their errors, as well as of the reduced $\chi^{2}$, for the combined fit of PAMELA data from 2006 to 2009 using both SPL and BPL scenarios. Two prominent features emerge. First, the best fit is always achieved with $\delta_{\odot}\sim1$, hinting at diffusion occurring close to the Bohm regime \cite{Berezinsky_book} in the solar system. Second, the solar mean-free-path normalization $\lambda_{0}$ tends to grow steadily from 2006 to 2009, which is in agreement with expectations since in those years the Sun was approaching the state of minimal activity, thus making diffusion faster. This is also in agreement with the general behavior found in \cite{Potgieter:2013cwj}. Remarkably, taking into account these effects, we are able to fit the data with very similar LIS for the different years. In particular, the fitted slopes are all compatible within errors in both scenarios. Inspection of the $\chi^{2}$ shows that the BPL model is slightly preferred over the SPL. 
%

\begin{table}[tbp]
\caption{Yearly variation of the best fit parameters, with their errors, in the SPL scenario. Errors on $\lambda_{0}$ and $\delta_{\odot}$ are $\pm0.01$ and $\pm0.05$ respectively and are set by the size of the grid we used to sample them.}
\begin{center}
\begin{tabular}{||c||c|c|c|c|c||}
\hline 
\multicolumn{6}{||c||}{Single power law} \\ 
\hline Year & $\lambda_{0}$ & $\delta_{\odot}$ &  $k_{0} \times 10^{-4}$ & $\alpha$ & $\chi^2/{\rm d.o.f.}$ \\
\hline 2006 & 0.18 & 1.0 & $1.74 \pm 0.02$ & $2.778 \pm 0.004$ & 9.99/87 \\
\hline 2007 & 0.20 & 1.0 & $1.82 \pm 0.02$ & $2.788 \pm 0.004$ & 8.30/87 \\
\hline 2008 & 0.22 & 1.0 & $1.75 \pm 0.02$ & $2.779 \pm 0.004$ & 10.5/87 \\
\hline 2009 & 0.24 & 0.9 & $1.89 \pm 0.04$ & $2.797 \pm 0.005$ & 6.79/87 \\
\hline
\end{tabular}
\end{center}
\label{tab:fitSP}
\end{table}%
\begin{table*}[tbp]
\caption{Yearly variation of the best fit parameters, with their errors, in the BPL scenario. Errors on $\lambda_{0}$ and $\delta_{\odot}$, not reported, are $\pm0.01$ and $\pm0.05$ respectively and are set by the size of the grid we used to sample them.}
\begin{center}
\begin{tabular}{||c||c|c|c|c|c|c|c||}
\hline 
\multicolumn{8}{||c||}{Broken power law} \\ 
\hline Year & $\lambda_{0}$ & $\delta_{\odot}$ &  $k_{0} \times 10^{-3}$ & $\alpha_1$ & $\alpha_2$ & $p_{\rm br}$  & $\chi^2/{\rm d.o.f.}$ \\
\hline 2006 & 0.20 & 0.8 & $3.29 \pm 0.59$ & $2.11 \pm 0.14$ & $2.790 \pm 0.005$ &  $1.85 \pm 0.12$ & 7.98/85 \\
\hline 2007 & 0.22 & 0.8 & $3.12 \pm 0.71$ & $2.35 \pm 0.10$ & $2.797 \pm 0.005$ &  $1.91 \pm 0.16$ & 6.41/85 \\
\hline 2008 & 0.22 & 0.9 & $4.08 \pm 0.31$ & $2.65 \pm 0.06$ & $2.792 \pm 0.004$ &  $1.72 \pm 0.05$ & 7.18/85 \\
\hline 2009 & 0.26 & 0.7 & $4.39 \pm 1.03$ & $2.32 \pm 0.13$ & $2.805 \pm 0.007$ &  $1.71 \pm 0.15$ & 5.90/85 \\
\hline
\end{tabular}
\end{center}
\label{tab:fitBP}
\end{table*}%


We then performed a joint fit of all the four PAMELA data sets fixing $\lambda_{0}$ and $\delta_{\odot}$ at their best fit values for each year, and assuming the same LIS model across the years (i.e.~with the same parameters for all the four years). The joined fits resulted into higher values of $\chi^2/{\rm d.o.f.}$ with respect to the individual fits, because of the drift of the $\lambda_{0}$ parameter with time. A better result was found assuming $\delta_{\odot}=1$ and $\lambda_{0}=0.18, ~0.20,~ 0.22,~ 0.24$ in the years 2006, 2007, 2008, 2009 for the SPL model, and taking $\delta_{\odot}=0.8$ and $\lambda_{0}=0.20,~ 0.22,~ 0.24, ~0.26$ in the years from 2006 to 2009 for the BPL case. 
We show in Figures~\ref{Fig3} and \ref{Fig4} the result of such a fit. The residuals exhibit very small fluctuations, within a few $\%$. The fit parameters for the SPL case are $k_{0}=(1.7781 \pm 0.0095) \times 10^4$ and $\alpha=2.783 \pm 0.002$ ($\chi^2/{\rm d.o.f.}=56.3/354$), while for the BPL case we have $k_{0}=(2837.95 \pm 355.90)$, $\alpha_1=2.365 \pm 0.047$, $\alpha_2=2.793 \pm 0.003$ and $p_{\rm br}=(1.96 \pm 0.09) \units{GeV}$ ($\chi^2/{\rm d.o.f.}=73.2/352$). In this case the BPL fit is slightly worse than the SPL, because the BPL model requires larger fluctuations of the solar parameters. It is worth to point out that the 2009 data show some large point-to-point fluctuations, probably due to the fact that these data were collected at the end of the XXIII Solar cycle.  

A comparable fit within the force-field model \cite{Gleeson_1968ApJ} fixing the LIS to one of our best fit models would yield larger $\chi^{2}$ and residuals of the order of 20\% for all years. Moreover, we find that the modulation potential steadily decreases in time from 0.6 to 0.4~GV, thus confirming that when approaching minimum solar activity, energy losses in the solar system are reduced.

Our LIS seems well compatible, within experimental uncertainties, with the one inferred from Fermi-LAT data \cite{Dermer:2013iwa,2012AIPC.1505...37C}. On the other hand, our LIS is very different from the one used in \cite{Potgieter:2013cwj}, which displays a quite strong flattening below 1~GeV. This explains the difference between our results and those shown in \cite{Potgieter:2013cwj}. Indeed, our parameter space scan included LIS spectra with such a flattening, but they turned out to provide poor fits of data for all years.

\begin{figure}[tbp]
\begin{center}
\includegraphics[width=1\columnwidth,keepaspectratio,clip]{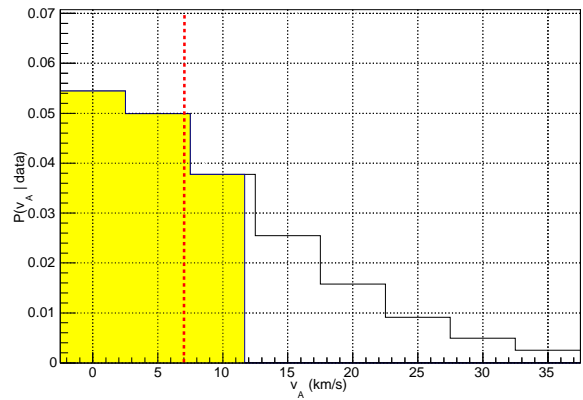}
\end{center}
\caption{Posterior probability distribution of $v_{A}$. The yellow region is allowed within 68\% CL. Mean value is indicated by the vertical dashed line.}
\label{fig:vA}
\end{figure}

We turn now to the question whether the inferred LIS can be compatible with galactic propagation models. For definiteness we show results for the BPL case (in the SPL case we obtain similar results but with an overall larger $\chi^{2}$). Within a single power-law {\em source injection} model we initially considered a very ample parameter space, including the parameters $D_{0}$, $\delta$, $\eta$ and $v_{A}$. However, the propagated proton spectrum is not sensitive to the normalization of the diffusion coefficient because it can be always readjusted by rescaling the spectrum normalization, so that $D_{0}$ plays no role. Also the parameter $\eta$ is not significantly constrained by this analysis, although an overall preference for $\eta<0$ is found (see \cite{Maurin:2010zp,DiBernardo:2009ku} for earlier studies of this possibilities). On the other hand, $\delta$ is degenerate with the unknown injection spectral index, therefore the proton spectrum is not sensitive to this parameter either. However, the parameter $v_{A}$ is significantly constrained by our analysis. We show in Fig.~\ref{fig:vA} the posterior probability distribution for the values of $v_{A}$, computed with the Bayesian Analysis Toolkit (BAT) \cite{BATweb,2009CoPhC.180.2197C}, marginalized over the whole parameter space we are considering, assuming for the moment no convection. The parameter $v_{A}$ is constrained to be less than $\sim12~\units{km/s}$ at 68\% CL.

If we allow also for convection, we obtain $v_{A}\lesssim 20~\units{km/s}$ and $dv_{C}/dz \lesssim25~\units{km/s/kpc}$. Allowing for a break in the injection spectrum yields comparable constraints, with best fit values being compatible with single power-law injection. We checked that our best-fit model also allows for a very good fit of the PAMELA Helium spectrum \cite{Adriani:2011cu} with the same slope at injection as for the proton spectrum.

{\em Conclusions:} The accurate proton data published by PAMELA and taken in different periods of solar activity offer now an unprecedented opportunity to investigate the effects of solar system propagation on the CR proton spectrum. We have exploited this possibility to infer for the first time the interstellar proton spectrum from these data by accounting also for the complexity of CR propagation in the solar system. Our technique provides useful complementary information to the proton LIS derived from $\gamma$-ray observations. A detailed comparison of the two techniques, which is left for future work, can help reducing experimental errors, as they are affected by very different systematics. 

Our CR proton LIS shows a slight preference for a single power-law over a broken power-law spectrum, that however produces a comparably good fit of data. Our spectrum seems well in agreement, within uncertainties, with the one inferred from Fermi-LAT $\gamma$-ray data. This is very reassuring and confirms the validity and complementarity of both techniques. Moreover, the proton and Helium LIS determined with our technique, together with electron and positron spectra constrained by the diffuse galactic synchrotron emission \cite{Strong:2011wd,DiBernardo:2012zu} and the new 3D propagation model developed to reproduce the recent AMS-02 results on the positron fraction \cite{Gaggero:2013rya}, can be taken as very well motivated CR spectra and distributions from which the diffuse galactic $\gamma$-ray emission can be computed.

We have also investigated the constraints placed by this LIS determination on galactic propagation models. First of all, our LIS is very well compatible with a single power-law injection. Indeed, assuming such an injection spectrum, we can place a strong constraint on the effectiveness of reacceleration, as $v_{A}$ is determined to be smaller than $12~\units{km/s}$ at 68\% CL. Strong reacceleration had been invoked to fit the downturn of secondary/primary nuclei ratios (namely, the B/C ratio) at about 1 GeV (see e.g.~\cite{Strong:1998pw,Trotta:2010mx}). A more recent analysis based on more advanced solar propagation models already showed however that the low energy part of the B/C ratio can be reproduced also with $v_{A}\sim10~\units{km/s}$ \cite{Maccione:2012cu}, in agreement with our findings. When convection is allowed for, the limit on $v_{A}$ is slightly weakened, and only convection effects are strongly constrained. This result is in qualitative agreement with the early claim made in \cite{DiBernardo:2009ku} on the basis of a statistical analysis of secondary nuclei, proton and antiproton spectra, and with more recent evidence coming for the study of the galactic diffuse synchrotron emission \cite{Strong:2011wd,DiBernardo:2012zu} and of the absolute positron spectrum \cite{DiBernardo:2012zu}. Although numerical comparisons are difficult because of different assumptions on the sizes of the diffusion regions, our results seem in agreement with the findings from semi-analytical codes \cite{Putze:2010zn}.

As a consistent picture of CR propagation is emerging from the combination of different data, we find these results very encouraging.  We expect that the wealth of new data that will be made available by the AMS-02 experiment \cite{ams02} will further improve our understanding of CR propagation.


{\em Acknowledgements:} We are grateful to Charles D. Dermer, Dario Grasso and Andrew W. Strong for fruitful discussions during the preparation of this manuscript and for their valuable contribution. LM acknowledges support from the Alexander von Humboldt foundation and partial support from the European Union FP7 ITN INVISIBLES (Marie Curie Actions, PITN-GA-2011-289442).
This work is based on the extensive use of the CPU-FARM of the Istituto Nazionale di Fisica Nucleare, Sezione di Bari.

\bibliographystyle{apsrev4-1}
\bibliography{LISHelio}

\end{document}